\documentstyle[preprint,prd,aps,eqsecnum,epsf]{revtex}

\begin{document}

\preprint{IF-UERJ 12/96\hspace{-26.6mm}\raisebox{-2.4ex}{March 1996}}

\title{Subcritical Fluctuations at the Electroweak Phase Transition}

\author{Rudnei O. Ramos\thanks{E-mail address: rudnei@vmesa.uerj.br}}

\address{{\it Universidade do Estado do Rio de Janeiro, }\\
{\it Instituto de F\'{\i}sica - Departamento de F\'{\i}sica Te\'orica,}\\
{\it 20550-013 Rio de Janeiro, RJ, Brazil}}

\maketitle

\begin{abstract}
\baselineskip 12pt
We study the importance of thermal fluctuations during the electroweak 
phase transition. We evaluate in detail the equilibrium number 
density of large amplitude subcritical fluctuations and discuss
the importance of phase mixing to the dynamics of the phase transition.
Our results show that, for realistic Higgs masses, the phase transition 
can be completed by the percolation of the true vacuum, induced by 
the presence of subcritical fluctuations.

\vspace{0.21in}

\noindent
PACS number(s): 98.80.Cq, 05.70.Fh.

\end{abstract}

\thispagestyle{empty}

\newpage

\setcounter{page}{1}

\section{Introduction}
The dynamics of the electroweak phase transition has received much attention
in recent years \cite{electroweak}.
The main reason for this interest is that, within the
context of big-bang cosmology, the standard electroweak model can,
in principle, satisfy the three conditions obtained by Sakharov for the
generation of the observed baryon asymmetry in the Universe \cite{Sakharov}:
(a) baryon number
violation;
(b) CP violation, and; (c) nonequilibrium dynamics. To date, most mechanisms
invoked to satisfy the third
condition make use of a first order phase transition. [For mechanisms based on
cosmic strings see Ref. \cite{Brandenberger}.] This is also true for
extensions of the standard model, which are currently favored by most authors,
due to difficulties in generating sufficient baryon number within the
minimal standard model \cite{Farrar}.

Based on estimates of the
quantum-corrected effective potential, nonequilibrium conditions are
generated by the motion of critical bubbles of the broken (true) phase,
which are
nucleated within the symmetric phase. For the several baryogenesis models which
rely on extensions of the standard model, the parameter space is large enough
to justify the assumption of a fairly strong first order transition.
However, analyses based on the 1-loop
finite
temperature effective potential for the minimal standard model\cite{GK,GG,GR}
and its improved
versions, show that the phase transition is very weak,
within the current lower bounds for the Higgs mass, $m_H \gtrsim
60 \;{\rm GeV}$ \cite{particle}, or even of second order for larger
Higgs
masses, $m_H \gtrsim 90 \; {\rm GeV}$ \cite{improved}. The weakness of the
transition is further supported by nonperturbative lattice computations
\cite{Kajantie}.

Our interest in the present paper is to further investigate \cite{GK,GKW} the
possible consequences of having a weak first order phase transition at the
electroweak scale. Although we will restrict our analysis to the standard
electroweak model, our results can be adapted to any of its extensions.
In fact, we will show that the strength of the transition can be used as a
new constraint on the parameters of the model, always a welcome addition to
the often large parameter space of extensions to the minimal standard model.

The interest in exploiting the dynamics of weak first order transitions goes
beyond its potential relevance for baryogenesis. As recent results have shown,
a sufficiently weak first order transition will exhibit a different dynamics
from the usual homogeneous nucleation results \cite{GH}; the extra free
energy available in large amplitude
fluctuations which are {\it not} included in the Gaussian
computation for the nucleation rate will act to decrease the decay barrier,
suppressing supercooling, and speeding up the completion of the phase
transition. For even weaker transitions, critical bubble nucleation may be
completely absent. Clearly, the information from the effective potential
is not sufficient to determine the details of the transition; mean-field
theory breaks down in the presence of large infrared corrections.

In order to quantify the above statements, we will make use of the subcritical
bubbles method \cite{GK,GKW}.
That is, we will model large amplitude thermal fluctuations
by Gaussian-shaped bubbles of approximately correlation volume. Previous
results based on a kinetic approach,
have indicated that such fluctuations can destroy the first-order
character of the transition for Higgs masses of order $m_H \gtrsim 55$ GeV
\cite{GG,GK}. Here we would like to complement this calculation by computing in
detail the nucleation rate for such configurations, which was previously
assumed for simplicity
to be $\Gamma = Am^4(T){\rm exp}[-B/T]$, where $A$ is a constant
of order unity, $m^4(T)$ is the curvature of the potential in the symmetric
phase, and $B$ is the free energy of the Gaussian configuration. Note that
this is a nontrivial exercise, as these configurations are not solutions to
the Euclidean equations of motion. Within reasonable approximations, we will
be able to obtain the equilibrium number density of these configurations
as a function of the tree-level Higgs mass, to show how the weakness of the
transition is closely related to the breakdown of the dilute gas
approximation.

The remainder of this paper is organized as follows. In Sec. II, we give
a brief review of subcritical bubbles in the context of the electroweak
standard model. In Sec. III, we compute the equilibrium number density of
subcritical fluctuations and discuss the choice of parameters in the
subcritical bubble configuration. There we also discuss the validity of
the approximations taken and the range of applicability of our results.
In Sec. IV, we discuss how the weakness of the
electroweak phase transition is related to the breakdown of the dilute gas
approximation; for large enough Higgs masses, subcritical bubbles percolate,
completing the transition.
In Sec. V, we have our main conclusions. Some
technical details are left for an Appendix.

\section{Subcritical Bubbles}

Following the work of Ref. \cite{GKW}, large amplitude fluctuations
describing thermal fluctuations are parameterized as,

\begin{equation}
\varphi_{sc}(r) = \varphi_A (T) \exp \left(-\frac{r^2}{R^2 (T)}
\right) \: ,
\label{sub}
\end{equation}

\noindent
where $\varphi_{sc} (r)$ describes (spherically symmetric) fluctuations
in the scalar field, with amplitude $\varphi_A$ and radius given by $R(T)$.
The minimum value for $R(T)$
should be compatible with the coarse-graining scale of the model.
Later, in section III, we will come back to the parameters in
(\ref{sub}) and discuss our choice for $\varphi_A$ and $R$.
The fluctuations described by (\ref{sub})  approximate rather well the
relevant field configurations, as recent work has shown \cite{GHK}.

\subsection{Free Energy for Electroweak Subcritical Bubbles}

Let us estimate the free energy associated with the configurations given
by (\ref{sub}). We shall use, as a particular case, the electroweak model.

In the electroweak model, for a Higgs self-coupling $\lambda \ll g^2$
(with $g$ denoting a generic gauge coupling), contributions due to
scalar loops to the finite temperature effective potential are small
compared to the contribution due to gauge fields and fermions. This
is a common approximation employed in the literature which results in a
1-loop finite temperature effective potential given by \cite{Hall}

\begin{equation}
V(\phi,T) = D (T^2 - T_2^2) \phi^2 - E T \phi^3 + \frac{\lambda_T}{4}
\phi^4 \:,
\label{V(T)}
\end{equation}

\noindent
where $D$ and $E$ are constants given in terms of the $W$ and $Z$ boson
masses and of the top quark mass as:

\begin{equation}
D = \frac{1}{24} \left[ 6 \left(\frac{m_W}{\sigma}\right)^2 +
3 \left(\frac{m_Z}{\sigma}\right)^2 + 6 \left(
\frac{m_t}{\sigma}\right)^2
\right] \simeq 0.169
\label{D}
\end{equation}

\noindent
and

\begin{equation}
E=\frac{1}{12 \pi} \left[ 6 \left(\frac{m_W}{\sigma}\right)^3 +
3 \left(\frac{m_Z}{\sigma}\right)^3 \right] \simeq
10^{-2} \:,
\label{E}
\end{equation}

\noindent
where $\sigma \simeq 246 \;{\rm GeV}$ is the vacuum expectation value
of the Higgs field. $m_W = 80.6 \; {\rm GeV}$, $m_Z = 91.2 \; {\rm
GeV}$ and we use $m_t \sim 174 {\rm GeV}$ \cite{top,particle}. $T_2$ in
$V(\phi,T)$ (the spinodal instability temperature) is
given by

\begin{equation}
T_2 = \sqrt{\frac{m_H^2 - 8 B \sigma^2}{4 D}} \: ,
\label{T2}
\end{equation}

\noindent
where $m_H^2 = (2 \lambda + 12 B) \sigma^2$ is the physical Higgs mass
and $B=\frac{1}{64 \pi^2 \sigma^4} (6 m_W^4 + 3 m_Z^4 - 12 m_t^4)
\simeq -0.00456$. $\lambda_T$ in $V(\phi,T)$ is the effective Higgs
self-coupling (at 1-loop) given by

\begin{equation}
\lambda_T = \lambda - \frac{1}{16 \pi^2} \left[
\sum_b g_b \left(\frac{m_b}{\sigma}\right)^4 \ln \left(
\frac{m_b^2}{c_b T^2} \right) - \sum_f g_f \left( \frac{m_f}{\sigma}
\right)^4 \ln \left(\frac{m_f^2}{c_f T^2} \right) \right] \: ,
\label{lambdaT}
\end{equation}

\noindent
where the sums are performed over bosons and fermions, with degrees of
freedom $g_b$ and $g_f$, respectively. In (\ref{lambdaT}), $\ln c_b =
5.41$ and $\ln c_f = 2.64$.

{}For temperatures $T< T_1$, where $T_1$ is given by the solution of
$E^2 T_1^2 = \frac{8}{9} D \left(T_1^2 - T_2^2\right) \lambda_{T_1}$,
$V(\phi,T)$ has minima at $\varphi =0$ and at

\begin{equation}
\varphi_+ (T) = \frac{1}{2 \lambda_T} \left[ 3 E T + \sqrt{9 E^2 T^2 -
8 D (T^2 - T_2^2) \lambda_T} \right] \: .
\label{true}
\end{equation}

At the critical temperature $T_c$,

\begin{equation}
T_c^2 = \frac{T_2^2}{1- \frac{E^2}{\lambda_T D}} \:,
\label{Tc}
\end{equation}

\noindent
we have $V(\phi=0,T_c) = V(\phi= \varphi_+,T_c)$ and below $T_c$,
$\phi= \varphi_f =0$ describes the metastable phase (the false vacuum),
while $\phi = \varphi_+ (T)$ is the stable phase (the true vacuum).

An important effect from higher loop corrections to $V(\phi,T)$ in
(\ref{V(T)}) is the reduction of the coefficient of the cubic term
$\phi^3$ by 2/3 \cite{Dine}. However, we are mostly interested in a
toy model
computation of the number density of subcritical fluctuations, and will
adopt the simplest 1-loop potential.
{}For recent considerations on improving
the 1-loop effective potential by resumming the most important infrared
contributions and higher order graphs see, for instance, Refs.
\cite{improved,Arnold,John}.

The free energy
for a given field configuration
$\varphi_{\rm sc} (T)$ is,

\begin{equation}
F(T) = \int d^3 x \left[\frac{1}{2} \left(\vec{\nabla}\varphi_{\rm sc}
\right)^2 + V(\varphi_{\rm sc},T) \right] \:.
\label{free}
\end{equation}

\noindent
{}From (\ref{sub}) and (\ref{V(T)}), we get for $F(T)$ the expression

\begin{equation}
F(T) = \alpha(\varphi_A) R(T) + \beta(\varphi_A) R^3 (T) \:,
\label{F(T)}
\end{equation}

\noindent
where

\[
\alpha(\varphi_A) = \frac{3 \sqrt{2}\pi^{\frac{3}{2}}}{8} \varphi_A^2
(T) \:,
\]
\begin{equation}
\label{ab}
\end{equation}
\[
\beta(\varphi_A) = \frac{\sqrt{2} \pi^{\frac{3}{2}}}{4} D (T^2 - T_2^2)
\varphi_A^2 (T) - \frac{\sqrt{3}\pi^{\frac{3}{2}}}{9} E T \varphi_A^3
(T) + \frac{\pi^{\frac{3}{2}}}{32} \lambda_T \varphi_A^4 (T) \:.
\]

\section{Number Density of Subcritical Bubbles}

\subsection{Partition Function for the Scalar Field in the
Electroweak Model}

Let us define in the Electroweak model the partition function

\begin{equation}
Z= \int D \phi D \chi_i e^{-\int_0^\beta d \tau \int d^3 x {\cal
L}_{\rm Eucl.} (\phi,\chi_i)} \:,
\label{Z}
\end{equation}

\noindent
where $\chi_i$ denotes gauge and fermions fields (and ghost fields) and
$\phi$ is the SU(2) doublet

\begin{equation}
\phi= \frac{1}{\sqrt{2}} \left(
\begin{array}{ll}
\phi_1 + i \phi_2 & \\
\phi_3 + i \phi_4 &
\end{array}
\right) \: ,
\end{equation}

\noindent
where $\phi_i$ ($i=1,\ldots,4$) are real scalar fields. The tree level
potential for the complex scalar field $\phi$, given by

\begin{equation}
V_0 (|\phi|) = -\mu^2 \phi^{\dag} \phi + \lambda (\phi^{\dag} \phi)^2 \:,
\label{V0}
\end{equation}

\noindent
for $\mu^2 > 0$, $\phi$ acquires a nonvanishing vacuum expectation
value $\langle |\phi| \rangle = \sigma$, which one assumes real and
along, for example, the real component $\phi_3$ of $\phi$. Thus, in the
broken phase, we define $\phi_3 ' = \phi_3 + \sigma$ and
$\phi_1,\phi_2$ and $\phi_4$ are the three Goldstone bosons.

Let us denote by $g_i$ the coupling of the field $\phi$ with the
$\chi_i$ fields. If $\lambda \ll g_i^2$, i.e., the interactions among
the $\phi$ field are weak compared with the $\phi-\chi_i$ interactions,
then we may formally integrate out the $\chi$ fields in (\ref{Z}) to
obtain

\begin{equation}
Z = \int D \phi e^{-W(\phi)}\:,
\label{ZW}
\end{equation}

\noindent
where

\begin{equation}
W(\phi) = - \ln \int D \chi_i e^{-\int_0^\beta d \tau \int d^3 x {\cal
L}_{\rm Eucl.} (\phi,\chi_i)} \: .
\label{W}
\end{equation}

\noindent
{}For vector fields, the integration measure above includes the gauge
fixing and ghost terms. We choose to work in the Landau gauge, which is
the one usually used in the studies of the electroweak phase transition.
Expanding $W(\phi)$ in a derivative expansion,

\begin{equation}
W(\phi) = \int_0^\beta d \tau \int d^3 x \left[ V_0 (|\phi|) +
V_\beta (|\phi|) + \hat{Z}(|\phi|) (\partial_\mu \phi)^{\dag}
(\partial^\mu \phi) + \ldots \right] \:,
\label{W2}
\end{equation}

\noindent
where $V_0(|\phi|)$ is the tree level potential (\ref{V0}) and $V_\beta
(|\phi|)$ is the contribution of the $\chi_i$ loops, coming from the
integration over the $\chi_i$ fields in (\ref{Z}), with the scalar field
$\phi$ in the external legs. Since $V(\phi,T)$, Eq. (\ref{V(T)}), is
obtained by neglecting scalar-boson contributions, which is analogous of
just making a functional integration on vector and fermion fields at the
1-loop approximation, $V_0 + V_\beta$ above, at the 1-loop approximation
for the $\chi_i$ fields, can be written as in Eq. (\ref{V(T)}).
$\hat{Z}(|\phi|)$ is the wave-function
renormalization factor. $\hat{Z}(|\phi|)$ has already been evaluated by
many authors, in the determination of an effective action for the Higgs
field with the objective to study corrections to critical bubbles
nucleation rates [see, for instance \cite{Buch2,Kripfganz,Walliser}],
where it is utilized a process of integration of fields, or field degrees of
freedom, analogous to the one done above to determine $W(\phi)$.
Although $\hat{Z}(|\phi|)$ receives nonvanishing 1-loop contributions,
it has also been shown in \cite{Buch2} that these contributions are
expected to yield only small corrections to the effective action and,
therefore, these corrections could be treated as perturbations. Later, in
section III.D, we show that the same approximation can be taken here,
at least for the range of temperatures and higgs masses we are
interested in. Therefore, as a first approximation,
we will neglect all wave function corrections and take
$\hat{Z}(|\phi|)\simeq 1$, for simplicity.
Thus, at 1-loop order, we write the ``effective'' action $W(\phi)$ as

\begin{equation}
W_1 (\phi) \simeq \int_0^\beta d \tau \int d^3 x \left[ | \partial_\mu
\phi|^2 + V(|\phi|,T) \right] \: .
\label{W1}
\end{equation}

\subsection{Number Density for Subcritical Fluctuations}

Well-known results show that, in a dilute gas approximation, the average
number of extended objects (for example, topological defects)
described by some field configuration $\varphi_c$
can be given by \cite{Langer67,Maki,Marques}

\begin{equation}
N_c = \frac{Z(\varphi_c)}{Z(\varphi_v)} \:,
\label{Nc}
\end{equation}

\noindent
where $Z(\varphi_c)~[Z(\varphi_v)]$
is the partition function of the system computed
by expanding the scalar field $\phi$ around the field (vacuum) configuration
$\varphi_c$ ($\varphi_v$).
{}For stable configurations the ratio in (\ref{Nc}) is real. However,
for unstable configurations (like, for example the sphaleron, the
critical bubble or bounce configuration and also for
$\varphi_{\rm sc}$, given by (\ref{sub})), the ratio in (\ref{Nc}) is
complex due to the existence of negative eigenvalues, associated with
the instability of the configuration. This is the case for subcritical
fluctuations. In this case, we will adopt the procedure of Arnold and
McLerran in \cite{McLerran} for the case of sphaleron configurations,
where they associated the average number of sphalerons to the total
rate of transitions multiplied by the time of a single transition,
giving, in the dilute gas approximation,

\begin{equation}
N_c \sim \Gamma \frac{2 \pi}{\omega_-} \sim {\rm Im}
\frac{Z(\varphi_c)}{Z(\varphi_v)}  \:,
\label{Im}
\end{equation}

\noindent
where $\Gamma$ is the transition rate given by $\Gamma \simeq
\frac{\omega_-}{\pi} {\rm Im} \frac{Z(\varphi_c)}{Z(\varphi_v)}$ and
$\omega_-$ is the negative eigenvalue. Note that in this case the final
result for (\ref{Im}) will not depend on the negative eigenvalue, as
shown in
\cite{McLerran}. Taking (\ref{Im}) as also valid for the case of
subcritical fluctuations, we can associate $\Gamma$ to the nucleation
rate of subcritical fluctuations and the possible negative eigenvalues,
which must appear in (\ref{Nc}), are the ones associated with the
collapse mode of the fluctuation. We will, therefore, evaluate $N_c$ for
subcritical fluctuations not taking into account the possible imaginary
eigenvalues associated with the instability of the configuration, that
is, we will adopt a procedure similar to the one expressed by (\ref{Im}).

Computing the partition functions will give us the
{\it equilibrium} values for relevant physical quantities.
In particular, the equilibrium number density
density of subcritical fluctuations, we will thus write as

\begin{equation}
n_{\rm sc} = \frac{N_{\rm sc}}{V} =\frac{1}{V} \frac{\int D \phi
e^{-W(\phi \rightarrow \varphi_{\rm sc}
+ \eta)}}{\int D \phi e^{-W(\phi \rightarrow \varphi_v + \zeta)}} \:,
\label{Nsc}
\end{equation}

\noindent
where $\eta \equiv \eta (\vec{x},\tau)$ and $\xi \equiv \xi
(\vec{x},\tau)$ are small perturbations around the configurations
$\varphi_{\rm sc}$ and $\varphi_v$ (which we take as the false vacuum
configuration $\varphi_f =0$), respectively. In the following section we
will discuss the limits of applicability of Eq. (\ref{Nsc}) within the
standard electroweak model and in other situations applicable to phase
transitions in general.

\subsection{1-loop Evaluation of $N_{\rm sc}$}

In the 1-loop approximation for $W(\phi)$ in (\ref{Nsc}), given by
(\ref{W1}), $W_1 (\phi)$ can be expanded about a field configuration
$\varphi_c$ as

\begin{eqnarray}
W_1 (\phi) &=& W_1 (\varphi_c) + \int d^4 x W_1 ' (\varphi_c;x) \eta(x)
+ \nonumber \\
&+& \frac{1}{2 !} \int d^4 x d^4 x' W_1 '' (\varphi_c;x,x') \eta(x)
\eta(x') + {\cal O} (\eta^3) \:,
\label{W1eta}
\end{eqnarray}

\noindent
where $\eta(x) = \phi(x) - \varphi(x)$ and $\int d^4 x = \int_0^\beta d
\tau \int d^3 x$. Taking $\eta(x)$ as small perturbations around the
field configurations $\varphi(x)$, the terms of order ${\cal O}(\eta^3)$
and higher can be treated as small perturbations. Note that for the
kind of configurations we are dealing with,
 $W_1 ' (\varphi) =
\frac{\delta W_1 (\phi)}{\delta \phi}|_{\phi=\varphi}$ does not vanish
in general ($\varphi_{\rm sc} (r)$ since is not a stationary solution of $W_1
(\phi)$.

Using (\ref{W1eta}) in (\ref{Nsc}) we can perform the gaussian
functional integrals and since we have already integrated out all other
fields interacting with the scalar field $\phi$, the scalar field
propagators will include loop (quantum) corrections from the other
fields coupled to $\phi$. We must, therefore, take some care when
performing the gaussian functional integral in $\phi$ in order to avoid
possible double-counting. {}From (\ref{W1eta}) and (\ref{Nsc}), with
$\varphi_v=\varphi_f=0$ and $W_1 ' (\varphi_v)=0$, we get

\begin{equation}
\frac{Z(\varphi_{\rm sc})}{Z(\varphi_v)} =
\left[ \frac{\det \bar{W}_1 '' (\varphi_{\rm sc})}{\det \bar{W}_1 ''
(\varphi_{v})} \right]^{- \frac{1}{2}} e^{-\Delta W_1} e^{W_{\rm sc} '
(\bar{W}_{\rm sc} '')^{-1} W_{\rm sc} '} \:,
\label{ZZ}
\end{equation}

\noindent
where $\Delta W_1 = W_1 (\varphi_{\rm sc}) - W_1 (\varphi_v)$ and
$\bar{W}_1 ''$ denotes the correct dressed (including the 1-loop quantum
corrections from the $\chi_i$ fields in (\ref{Z})) inverse propagator
for the $\phi$ field:

\begin{equation}
\bar{W}_1 '' (\varphi) = -\Box + m_\beta^2 (\varphi) \:,
\label{prop}
\end{equation}

\noindent
with $m_\beta^2 (\varphi) = 2 D (T^2 - T_2^2) + 3 \lambda_T \varphi^2$.
Note that $m_\beta (\varphi)$ is the classical Higgs mass corrected by
the self-energy ($T\neq 0$) corrections coming from the $\chi_i$ fields
(fermions and gauge bosons). $m_\beta^2 (\varphi)$ does not include the
term proportional to $E$ of (\ref{V(T)}), which would give origin to a
linear term in (\ref{prop}). This term is absent since it would not
appear in the self-energy corrections to the scalar field $\phi$ and
also that the presence of such a term in (\ref{prop})
is well known to
lead to a wrong counting of loop corrections to the
scalar field $\phi$ effective potential (see for instance,
ref. \cite{Dine}).

In (\ref{ZZ}) we also have that

\begin{equation}
W_{\rm sc} ' (\bar{W}_{\rm sc} '')^{-1} W_{\rm sc} ' =
\int d^4 x d^4 x' W_1 ' (\varphi_{\rm sc};x) W_1 ' (\varphi_{\rm sc};x')
\langle x| \left[ \bar{W}_1 '' (\varphi_{\rm sc};x,x') \right]^{-1} |x'
\rangle \:,
\label{WWW}
\end{equation}

\noindent
where

\begin{equation}
\langle x| \left[ \bar{W}_1 '' (\varphi_{\rm sc};x,x') \right]^{-1} |x'
\rangle = \frac{1}{\beta} \sum_{n=-\infty}^{+\infty} \int
\frac{d^3 k}{(2 \pi)^3} \frac{e^{i \omega_n (\tau-\tau') +
i \vec{k}.(\vec{x} -
\vec{x}')}}{\omega_n^2 + \vec{k}^2 + m_\beta^2 (\varphi_{\rm sc})} \:,
\label{xWx}
\end{equation}

\noindent
where $\omega_n = \frac{2 \pi n}{\beta}$ ($n=0,\pm 1, \pm 2,\ldots$)
are the Matsubara frequencies. In appendix A we solve (\ref{WWW})
explicitly.

In order to compute the determinant ratio in (\ref{ZZ}), we must first
isolate the possible zero modes in $\det \bar{W}_1 '' (\varphi_{\rm
sc})= \det [ -\Box + m_\beta^2 (\varphi_{\rm sc})]$.
Note that, as exposed above, here we will not take into account possible
imaginary eigenvalues.
{}Noting
that $W_1 (\phi)$, Eq. (\ref{W1}), and the corresponding field
equation, $\frac{\delta W_1}{\delta \phi}=0$, are translational
invariant. Therefore, even though if $\varphi_{\rm sc}$ is not a
solution of the field equation, there must be three zero eigenvalues
associate to $\varphi_{\rm sc}$, related to the three translational
modes. We also have three more zero modes that are related to the
rotational symmetry $SU(2)$, associated with the three Goldstone bosons
in the broken phase. We handle these zero modes by the standard way, by
introducing collective coordinates, such that the determinantal ratio in
(\ref{ZZ}) can be written as:

\begin{equation}
\left[\frac{\det \bar{W}_1 '' (\varphi_{\rm sc})}{\det \bar{W}_1 ''
(\varphi_v)} \right]^{-\frac{1}{2}}  = \Omega_{\rm trans.} \Omega_{\rm
rot.} \left[\frac{\det' \bar{W}_1 '' (\varphi_{\rm sc})}{\det \bar{W}_1 ''
(\varphi_v)} \right]^{-\frac{1}{2}} \:,
\label{det}
\end{equation}

\noindent
where $\Omega_{\rm trans.}$ is the usual factor coming from the
translational modes, given by

\begin{equation}
\Omega_{\rm trans.} = \left[\frac{\Delta W_1}{2
\pi}\right]^{\frac{3}{2}} V
\label{wtrans}
\end{equation}

\noindent
and $\Omega_{\rm rot}$, due to the rotational modes, has been
explicitly obtained in \cite{Buch2}, for the case of fluctuations
around the critical bubble configuration, and in our case we can write
the analogous expression for the $\varphi_{\rm sc}$ configuration as

\begin{equation}
\Omega_{\rm rot.} = \frac{\pi^2}{2} \left[ \frac{\beta}{2 \pi}  \int
d^3 x \varphi_{\rm sc}^2 (\vec x) \right]^{\frac{3}{2}} \:,
\label{wrot}
\end{equation}

\noindent
In (\ref{det}) the
prime in the determinant is to indicate that the six zero eigenvalues
have been excluded. In (\ref{wtrans}) $V$ is the space volume.

{}From (\ref{W1}), we can write the determinantal ratio in (\ref{ZZ}),
in momentum space, as

\begin{equation}
\left[\frac{\det \bar{W}_1 '' (\varphi_{\rm sc})}{\det \bar{W}_1 ''
(\varphi_v)} \right]^{-\frac{1}{2}}  = \exp \left\{ - \frac{1}{2} \ln
\left[ \frac{\prod_{n =-\infty}^{+\infty} \prod_i (\omega_n^2 + E_i^2
(\varphi_{\rm sc}))}{\prod_{l=-\infty}^{+\infty} \prod_j (\omega_l^2 +
E_j^2 (\varphi_v))} \right] \right\}\:,
\label{detdet}
\end{equation}

\noindent
where the productories in $i$ and $j$ are, formally, over the
eigenvalues $E_i^2 (\varphi_{\rm sc})$ and $E_j^2 (\varphi_{v})$,
respectively. In (\ref{detdet}), we have used the identity $\ln \det
\hat{O} = {\rm tr} \ln \hat{O}$. Taking into account the six zeroes modes of
(\ref{detdet}), associated with the three Goldstone bosons of the Higgs
doublet, in the broken phase, and the three translational modes, we
obtain

\begin{equation}
\left[\frac{\det \bar{W}_1 '' (\varphi_{\rm sc})}{\det \bar{W}_1 ''
(\varphi_v)} \right]^{-\frac{1}{2}}= \Omega_{\rm trans.} \Omega_{\rm
rot.} \exp\left\{ - \frac{1}{2} \ln \left[
\frac{\left(\prod_{n=1}^{+\infty} \omega_n^2\right)^{12} \prod_i '
\prod_{n=-\infty}^{+\infty} (\omega_n^2 + E_i^2 (\varphi_{\rm sc}))}{
\prod_j \prod_{l=-\infty}^{+\infty} (\omega_l^2 + E_j^2 (\varphi_v))}
\right] \right\} \:,
\label{det2}
\end{equation}

\noindent
where the factors $\Omega_{\rm trans.}$ and $\Omega_{\rm rot.}$ are
given by (\ref{wtrans}) and (\ref{wrot}), respectively. The prime in
$\prod_i$ in (\ref{det2}) is a reminder that those zero modes,
associated with (\ref{wtrans}) and (\ref{wrot}) have been excluded from
the product of eigenvalues.

Using the identity:

\begin{equation}
\prod_{n=1}^{+\infty} \left( 1 + \frac{a^2}{(2 \pi n)^2} \right) =
\frac{\sinh (a/2)}{a/2} \:,
\label{identity}
\end{equation}

\noindent
we get for (\ref{det2}) the expression

\begin{eqnarray}
\left[\frac{\det \bar{W}_1 '' (\varphi_{\rm sc})}{\det \bar{W}_1 ''
(\varphi_v)} \right]^{-\frac{1}{2}} &=& \Omega_{\rm trans.} \Omega_{\rm
rot.} \exp\left\{ - \frac{1}{2} \left[12 \ln \beta + 2 \sum_i '
\left[\frac{\beta}{2} E_i (\varphi_{\rm sc}) + \ln \left(1-e^{-\beta E_i
(\varphi{\rm sc})}\right) \right] - \right. \right.\nonumber \\
&-& \left.\left. 2 \sum_j
\left[\frac{\beta}{2} E_j (\varphi_{v}) + \ln \left(1-e^{-\beta E_j
(\varphi{v})}\right) \right] \right] \right\} \:.
\label{detsum}
\end{eqnarray}

{}For the field configuration $\varphi_{\rm sc}$, given by (\ref{sub}),
and the vacuum configuration $\varphi_v$ (the false vacuum,
$\varphi_v=0$), from (\ref{W1}), we can write that

\begin{eqnarray}
\hspace{-3cm}\sum_{(\rm continuum)} \left[\frac{\beta}{2}
E (\varphi) + \ln
\left(1-e^{-\beta E(\varphi)} \right) \right] & \simeq  & \int d^3 x
\int \frac{d^3 k}{(2 \pi)^3} \left[ \frac{\beta}{2} \sqrt{\vec{k}^2 +
m_\beta^2 (\varphi)} \; + \right. \nonumber \\
& + & \left.
\ln \left(1-e^{-\beta \sqrt{\vec{k}^2 + m_\beta^2
(\varphi)}}\right) \right] \:,
\label{corr}
\end{eqnarray}

\noindent
where $m_\beta^2 (\varphi) = 2D (T^2 - T_2^2) + 3 \lambda_T \varphi^2$.
Note that the terms like $\int d^3 k \sqrt{\vec{k}^2 + m^2}$ in
(\ref{corr}), that are ultraviolet divergent, can be subtracted by
introducing the usual counterterms of renormalization for the scalar
field loops, rendering the exponent in (\ref{detsum}) finite.

Using (\ref{corr}) in (\ref{detsum}), Eq. (\ref{ZZ}) can therefore be
written as

\begin{equation}
\frac{Z(\varphi_{\rm sc})}{Z(\varphi_v)} \stackrel{\rm 1-loop \;
approx.}{\simeq} \Omega_{\rm trans.} \Omega_{\rm rot.} T^6 \exp \left[-
\frac{\Delta F_{\rm eff}(T)}{T} \right] \:,
\label{ZZ2}
\end{equation}

\noindent
where $\Delta F_{\rm eff}(T)$ denotes an effective free energy for
subcritical bubbles, given by

\begin{eqnarray}
\Delta F_{\rm eff}(T) &=& F(T) + \int d^3 x \left[V_\phi (\varphi_{\rm
sc},T) - V_\phi (\varphi_v,T) \right] - \nonumber \\
&-& T \int d^4x d^4 x' W_1 ' (\varphi_{\rm sc};\vec x) W_1 '
(\varphi_{\rm sc};\vec{x} ') \langle x | \left[ \bar{W}_1 ''
(\varphi_{\rm sc};x,x') \right]^{-1} | x' \rangle \:,
\label{DF}
\end{eqnarray}

\noindent
where $F(T)$ is given by (\ref{F(T)}). The scalar field quantum
contribution in (\ref{DF}), in the high T limit, is given by

\begin{eqnarray}
\int d^3 x \left[ V_\phi (\varphi_{\rm sc},T) - V_\phi (\varphi_v,T)
\right] &\stackrel{\rm high \; T}{\simeq}& 4 \pi \int_0^\infty d r r^2
\frac{T^2}{24} 3 \lambda_T \varphi_{\rm sc} (r) = \nonumber \\
&=& \frac{\sqrt{2} \pi^{\frac{3}{2}}}{32} \lambda_T
\varphi_A^2 (T) T^2 R^3
(T) \:,
\label{term2}
\end{eqnarray}

The last term in
(\ref{DF}), from (\ref{WWW}) and (\ref{xWx}), can be written as

\begin{equation}
\int d^3 x d^3 x ' \left[ - \vec{\nabla}^2 \varphi_{\rm sc} (\vec x) +
V'(\varphi_{\rm sc}(\vec x),T) \right] \left[ - \vec{\nabla'}^2
\varphi_{\rm sc} (\vec{x} ') + V ' (\varphi_{\rm sc}(\vec{x} '),T)
\right]\; I (|\vec x - \vec x '|) \:,
\label{term}
\end{equation}

\noindent
where $I(|\vec x - \vec{x}'|)$, in the high temperature limit,
is given by (see the Appendix)

\begin{equation}
I(|\vec x - \vec{x} '|) \stackrel{\rm high \: T}{\simeq} \frac{\pi}{(2
\pi)^2 | \vec x - \vec{x} ' | T} e^{- m_\beta |\vec{x} - \vec{x} '|}
\:.
\label{I1}
\end{equation}

Using the above expression in (\ref{term}), we can see that the largest
contribution will come for values of $|\vec{x} - \vec{x}'|$ close to
the inverse of $m_\beta$, i.e., for values close to the correlation
length $\xi(T)$ ($\sim 1/m_\beta$). We therefore may restrict $|\vec{x}
- \vec{x}'|$ to the size of the subcritical fluctuations and we here
consider $R(T) \sim
\xi(T)$), thus
obtaining

\begin{equation}
I(|\vec{x}-\vec{x}'|) \sim \frac{\pi}{(2 \pi)^2 T R(T) e} \:.
\label{I2}
\end{equation}

Substituting $\varphi_{\rm sc}(\vec{x})$ in (\ref{term}) and using the above
expression for $I(|\vec x - \vec x '|)$, we get

\begin{equation}
W' (\bar{W} '')^{-1} W' \simeq \frac{\pi^2 \varphi_A^2 (T)
R^5 (T)}{T e}
\left[ D(T^2 - T_2^2) - 3 \frac{\sqrt{2}}{8} ET \varphi_A (T)  +
\frac{\sqrt{3}}{18}  \lambda_T \varphi_A^2 (T) \right]^2 \:.
\label{WWW2}
\end{equation}

\noindent
Using the expression for $F(T)$, Eq. (\ref{F(T)}), and Eqs.
(\ref{term2}) and (\ref{WWW2}), we get for $\frac{Z(\varphi_{\rm
sc})}{Z(\varphi_v)}$, in the high temperature limit and at 1-loop order,
the expression

\begin{equation}
\frac{Z(\varphi_{\rm sc})}{Z(\varphi_v)} = V T^3 A(T) \exp [- B(T)] \:,
\label{ZAB}
\end{equation}

\noindent
where $A(T)$ and $B(T)$ are given by

\begin{eqnarray}
A(T) \stackrel{\rm 1-loop, \: high \: T}{\simeq}
R^{6}(T) \varphi_A^6 (T)  \frac{\pi^{\frac{7}{2}}}{16}
 \left[ \frac{3 \sqrt{2}}{8} \right. &+& \left.
\frac{\sqrt{2}}{4} D (T^2 -T_2^2) R^2
(T) - \frac{\sqrt{3}}{9} ET \varphi_A (T) R^2 (T) + \right.
\nonumber \\
& + & \left. \frac{\lambda_T}{32} \varphi_A^2 (T)
R^2 (T) \right]^\frac{3}{2}
\label{A}
\end{eqnarray}

\noindent
and

\begin{equation}
B(T)  \stackrel{\rm 1-loop, \: high \: T}{\simeq}  \frac{\Delta F_{\rm
eff} (T)}{T} \:,
\label{B}
\end{equation}

\noindent
with $\Delta F_{\rm eff} (T)$ given by (\ref{DF}) together with Eqs.
(\ref{F(T)}), (\ref{term2}) and (\ref{WWW2}).

{}From (\ref{Nc}), the equilibrium number density of
subcritical fluctuations, in the dilute gas approximation, is given by

\begin{equation}
n_{\rm sc} (T) = \frac{1}{V} \frac{Z(\varphi_{\rm sc})}{Z(\varphi_v)}
\:.
\label{nsc}
\end{equation}

\subsection{A Discussion of the Validity of the Approximations and the
Choice for the Parameters $\varphi_A$ and $R$}

We expect that the expression for $n_{\rm sc}$, obtained in the dilute
gas approximation, be
valid as long as the following holds

\begin{equation}
\left[ n_{\rm sc} (T) \right]^{\frac{1}{3}} R (T) \ll 1 \:,
\label{val}
\end{equation}

\noindent
We can also express the
condition in terms of the volume occupied by the fluctuations described
by $\varphi_{\rm sc}$, at some fixed temperature $T$, $V_{\rm sc}(T)$,

\begin{equation}
V_{\rm sc} (T) = \frac{4 \pi}{3} R^3 (T) n_{\rm sc} (T) V \:,
\label{Vsc}
\end{equation}

\noindent
such that (\ref{val}) can be reexpressed as

\begin{equation}
\frac{V_{\rm sc} (T)}{V} \ll 1
\:,
\label{val2}
\end{equation}

\noindent
that is, the dilute gas approximation is a good approximation as long
as the volume occupied by the subcritical fluctuations be small.

Another approximation that we considered, the neglected of wave
function corrections to the effective action, can be thought to be
somewhat problematic. However, we can estimate their relative
contribution to our results and inquiry whether our approximation of
neglecting these contributions is good enough. {}For such estimation we
may take the leading order correction to $\Delta F_{\rm eff} (T)$, due to
the wave function correction coming from the functional integration
over the gauge boson fields. As an order of estimative, we may take
the result obtained in \cite{Buch2}, for the SU(2)-Higgs model,
which we write as,
in neglecting contributions due to plasma masses,

\begin{equation}
{\cal Z}(\varphi,T) = \frac{11}{32 \pi} \frac{g T}{\varphi} \:,
\label{wave}
\end{equation}

\noindent
where $g = 2 m_W/\sigma \simeq 0.328$. An estimate of the relative
contribution of the above factor can be given by the ratio (as in
\cite{Buch2})

\begin{equation}
\delta = \frac{\int d^3 x {\cal Z}(\varphi_{\rm sc},T) \left(\vec \nabla
\varphi_{\rm sc}\right)^2}{\int d^3 x \left(\vec \nabla
\varphi_{\rm sc}\right)^2} = \frac{11 g}{4 \sqrt{2} \pi}
\frac{T}{\varphi_A} \:,
\label{deltaZ}
\end{equation}

\noindent
where we have used (\ref{sub}) in the rhs of the above equation. It is interesting
to note that the above expression do not depend on the
subcritical bubble configuration radius $R$.

The range of applicability of the dilute gas approximation and the
validity of the approximation of neglecting wave function contributions
is determined once the parameters $\varphi_A$ and $R$, of our {\it ansatz},
Eq. (\ref{sub}), are set.

In the following, we take the amplitude $\varphi_A$
as been given by the temperature dependent broken minimum, $\varphi_+
(T)$, Eq. (\ref{true}). {}For $R$ we assume it as been given by the
correlation length $\xi(T)$, given by the inverse of the temperature
dependent mass in $V(\phi,T)$. We next justify these
assumptions. We expect that fluctuations in the scalar field
would appear in the system with arbitrary amplitudes and sizes.
However, physically, we can set some limits on these parameters. {}For
example, fluctuations with too small amplitudes are already summed over
in the computation of the (coarse-grained) effective potential. The
relevant fluctuations for the dynamics of the phase transition can
roughly be identified \cite{GH,GHK} as been those with amplitudes
$\varphi_A \gtrsim \varphi_{\rm max}$, where $\varphi_{\rm max}$ is the
value of the higgs field at the maximum of the effective potential.
Among these fluctuations in the false vacuum, those with $\varphi_A$
close to the local minimum $\varphi_+$ are certainly the most expected.
About the fluctuations radius, $R$, their minimum radius must be
compatible with the coarse-graining scale of the model and the
coarse-grained effective potential. In general, we can expect that $R
\gtrsim \xi (T)$. Here, we implicitly assume that the most probable
fluctuations in the system are those with radius close to the
correlation length $\xi(T)$. A better approximation would be taking $R$
as the average radius of the fluctuations. Unfortunately, a reliable
method to determine this quantity is still lacking. Recently, the
authors of ref. \cite{mori} propose a method to determine
the average radius $\bar{R}$ of fluctuations, considering $R$ as a
truly dynamical variable, however, the obtained result for $\bar{R}$ is
much smaller than $\xi(T)$. In another recent method proposed by the
authors in \cite{GH,GHK}, a statistical mechanical method is used for
studying the importance of phase mixing in the phase transition, where
the volume fraction occupied by fluctuations is determined by summing
over all possible fluctuations of different sizes and amplitudes. In
the next section, by computing this volume fraction within our
approximations, we will be able to compare our results with the ones
obtained in the statistical mechanics method.

With $\varphi_A=\varphi_+(T)$ and $R=\xi(T)$, we can immediately evaluate
(\ref{deltaZ}) and have an estimative of the leading wave function
contribution. In Figure 1 we have given $\delta$ as a function of the
higgs mass, with $\delta$ computed at the critical temperature $T_c$.
Our result for the fraction given by (\ref{deltaZ}) is much like the
same one as given in \cite{Buch2}, where $\delta \geq 1$ for
$m_H \gtrsim 80 GeV$. However, for such larger higgs masses it is well
known that the perturbative expansion breaks down and the effective
potential becomes unreliable \cite{GR,improved,Buch}. {}For the
interval of higgs masses we will be interested in, the wave function
correction is small and, if higher order corrections are taken into
account and plasma masses are included in ${\cal Z}(\phi,T)$, we expect
the wave function contributions be even smaller, as shown in
\cite{Buch2}.

\section{Percolation of Subcritical Fluctuations}

Studies of the electroweak phase transition, through the analyses of
the effective potential \cite{Buch} and direct numerical simulations
\cite{Gleiser}, indicate that the nature of the electroweak phase
transition is most possible of first order, but it has a excitation
barrier to small, qualifying the phase transition as very weakly first
order. In these situations it has been argued that the dynamics of the
phase transition could  be quite different from the usual process of
critical bubble nucleation. In fact, due the weak nature of the phase
transition, (subcritical) fluctuations over the barrier can become the
main mechanism by which the phase transition can complete. Here, we
analyze the importance of a possible large amount of phase mixing at
temperatures higher than the nucleation temperature of critical
bubbles. In this sense, we argue that if the volume of true vacuum
phase fluctuations becomes higher than a certain value, percolation of
the fluctuations can occur. We thus have
a possible scenario where the
phase transition can be completed just by those over the barrier
subcritical fluctuations.

A well known result of statistical mechanics of dynamics of cluster
systems \cite{Staufer} show that there is a critical probability value
$p_c$, with the cluster probability defined by

\begin{equation}
p = \frac{\rm Cluster \; Volume}{\rm System \; Volume} \:,
\label{perc}
\end{equation}

\noindent
where, in three
space dimensions, this critical percolation probability is roughly
$p_c \sim 0.3$
\cite{Staufer} and
beyond which clusters are favored
to coalesce and grow, filling the whole volume of the system.

{}From our definition of the volume occupied by the field
configurations $\varphi_{\rm sc}$, at a fixed temperature $T$, Eq.
(\ref{Vsc}), it is quite natural to define, therefore, a percolation
temperature $T_p$
as given by

\begin{equation}
\frac{V_{\rm sc} (T_p)}{V} \sim 0.3 \:.
\label{Vperc}
\end{equation}

\noindent
{}From Eq. (\ref{Vsc}), we obtain

\begin{equation}
\frac{4 \pi}{3} R^3 (T_p) n_{\rm sc} (T_p) \simeq 0.3 \:.
\label{Vperc2}
\end{equation}

\noindent
Condition (\ref{Vperc2}) gives us the temperature (as a function of the
Higgs mass) below which subcritical fluctuations become favorable to
coalesce and grow, forming large regions of the true vacuum phase.
If $T_p$ is larger than the nucleation temperature $T_N$, that marks
the onset of critical bubbles nucleation, then we have an effective
mechanism by which the phase transition can complete, as explained
above, only due the dynamics of subcritical fluctuations of the stable
phase $\varphi_+$, inside the false vacuum phase. Note that at $T_p$,
condition (\ref{Vperc}), due to  (\ref{val2}), will also signal
the breakdown of the dilute gas approximation. Thus, for temperatures
too close to $T_p$, the obtained expression for $n_{\rm sc}$, Eq.
(\ref{nsc}), would just give a qualitative indication
of the importance of subcritical fluctuations
during the phase transition.
We must, however, comment on the alternative method of the authors of
refs. \cite{GH,GHK}, where a statistical mechanical approach is used
for studying the dynamics of subcritical bubbles, modeled there as in
Eq. (\ref{sub}). By constructing a kinetic equation for the (dynamical)
number density of fluctuations, they could obtain an analytical
expression for $V_{\rm sc}/V$, within appropriate approximations. Their
kinetic approach also break down for $V_{\rm sc}/V$ close to the
critical percolation probability, where fluctuations become more dense
and the treatment of the coalescence of fluctuations becomes important.
Besides the different approach, the method applied to the electroweak
phase transition (see Gleiser's talk in \cite{GHK}) show that $V_{\rm
sc}/V$ changes sharply for $m_H \gtrsim 60 GeV$, just as the result
obtained here for $V_{\rm sc}/V$ shows (Figure 2).
Also, recent numerical simulations
of very weak first order phase transitions \cite{Gleiser} give support
to our
results. All these dynamical studies
indicate the validity of the various
approximations here adopted in order to make possible to arrive to an
analytical
expression for $n_{\rm sc}$.

In our analyses it is enough to study the ratio $V_{\rm sc}/V$ at the
critical temperature $T_c$, for which we get the plot shown in Figure
2, in terms of the Higgs mass $m_H$, where we have used the
approximation for $\lambda_T$,

\begin{equation}
\lambda_T \sim \lambda \sim 0.08 \left(\frac{m_H}{100 {\rm GeV}}
\right)^2 \:.
\end{equation}

{}From Figure 2 we see that for a Higgs mass of $\sim 60 {\rm GeV}$,
our results indicate that phase mixing is so large that
the volume of subcritical fluctuations in the broken phase
start becoming relevant for the
dynamics of the phase transition. However, due to the condition of
validity of the dilute gas approximation used in our evaluation,
Eq. (\ref{val2}), our results can only be interpreted
qualitatively for $m_H \gtrsim 60 GeV$. Besides these limitations,
the results
obtained here are in accordance to recent results based on dynamical
studies of very weak phase transitions, as discussed above.

The main result of this paper is the computation of the
equilibrium density for the field configuration $\varphi_{\rm sc}$,
with the full evaluation of the preexponential term $A(T)$ given by
(\ref{A}) and the effective free energy $\Delta F(T)$, Eq. (\ref{DF}).
Previous results have given $n_{\rm sc}$ just in terms of a Boltzman
distribution form, with free energy given by (\ref{F(T)}) and prefactor
given by just $T^3$.
{}From (\ref{ZAB}) and (\ref{nsc}) we see that the preexponential
factors differ by the factor $A(T)$, which we show in Figure 3, as
a function of the Higgs mass, which shows that the prefactor can assume
large values for small values of the higgs mass. In Figure 4, we also
compare the exponential factor $\Delta F_{\rm eff} (T)$ with $F(T)$,
both compute at $T=T_c$.

\section{Conclusions}

We have explicitly evaluated the equilibrium density of subcritical
fluctuations and shown that the same can differ substantially from the
usual expression. We have also analyzed the relative importance
of subcritical fluctuations during the electroweak phase transition,
showing that the mechanism for completion of the phase transition can be
quite different from the usual critical bubble nucleation
in a first order phase transition. For values of the Higgs mass at and
higher than the experimental lower bound of $\sim 60 {\rm GeV}$, our results
show that substantial phase mixing can be present prior to critical bubble
nucleation, changing the dynamics of the phase transition.
Our results for the number density and volume occupied
by those subcritical fluctuations can be interpreted as an average
value at some fixed temperature, since we expect that
in some large volume $V \gg R^3 (T)$ ($R(T) \sim \xi (T)$) we have both
processes of nucleation of subcritical bubbles and their
shrinking,
keeping their average density, at that temperature, constant in that
large volume.

We must also remember that the results obtained here are in accordance
with other studies involving the relevance of thermal fluctuations
during the electroweak phase transition \cite{GR} and recent studies
of the dynamics
of very weak first order phase transitions, both by numerical
simulations as by a statistical mechanics approach for the dynamics of
fluctuations modeled by $\varphi_{\rm sc}$, Eq. (\ref{sub}).

\vskip 1.cm

\centerline{\large \bf Acknowledgements}

\vskip 0.5cm
I would like to thank Marcelo Gleiser for the many discussions we have
had on the
subject of the paper and for his valuable suggestions, which improved
the paper considerable.
This work was supported in part by
Conselho Nacional de Desenvolvimento Cient\'{\i}fico e Tecnol\'ogico
(CNPq-Brazil).

\newpage

\appendix

\section{Evaluation of Eq. (3.28)}

In this appendix we evaluate (\ref{WWW}) and get the expression for
$I(|\vec x - \vec{x}'|)$, Eq. (\ref{I1}). The term expressed in
(\ref{WWW}),

\begin{equation}
W_{\rm sc} ' ( \bar{W}_{\rm sc} '')^{-1} W_{\rm sc} ' =
\int d^4 x d^4 x' W_1 ' (\varphi_{\rm sc};x) W_1 ' (\varphi_{\rm
sc};x')
\langle x | \left[ \bar{W}_1 '' (\varphi_{\rm sc};x,x') \right]^{-1} |
x' \rangle \:,
\label{B1}
\end{equation}

\noindent
where $W_1 ' = \frac{\delta W_1}{\delta \phi}$, is obtained from
Eq. (\ref{W1}). Since $W_1 ' (\phi)$ computed at $\phi=\varphi_{\rm sc}
(\vec x)$ is independent of the Euclidean time, the two Euclidean time
integrations in (\ref{B1}) applied directly to $\langle x | \left[
\bar{W}_1 ''
(\varphi_{\rm sc};x,x') \right]^{-1} |x' \rangle$ which we called
$I(|\vec x - \vec{x}'|)$:

\begin{equation}
I(|\vec x - \vec x '|) = \int_0^\beta d \tau \int_0^\beta d \tau '
\langle x | \left[ \bar{W}_1 '' (\varphi_{\rm sc};x,x') \right]^{-1} |
x' \rangle \:.
\label{B2}
\end{equation}

\noindent
Using (\ref{xWx}), performing the sum in the Matsubara frequencies in
(\ref{xWx}) and using the result in (\ref{B2}), we can perform the two
Euclidean time integrations in (\ref{B2}) in order to get the final
result

\begin{equation}
I(|\vec x - \vec x '|) = \int \frac{d^3 k}{(2 \pi)^3} \frac{e^{i \vec k
. (\vec x - \vec x ')}}{2 \left(\vec{k}^2 + m_\beta^2
\right)^{\frac{3}{2}}}
\sinh \left( \beta \sqrt{ \vec{k}^2 + m_\beta^2} \right) \:.
\label{B3}
\end{equation}

\noindent
Performing the angular integrations and the $k$ integration of
(\ref{B3}), we get, in the high temperature limit, $ \beta m_\beta
\ll 1$, the result given in Eq. (\ref{I2}).

\newpage
\epsfysize=20cm
\centerline{\epsfbox{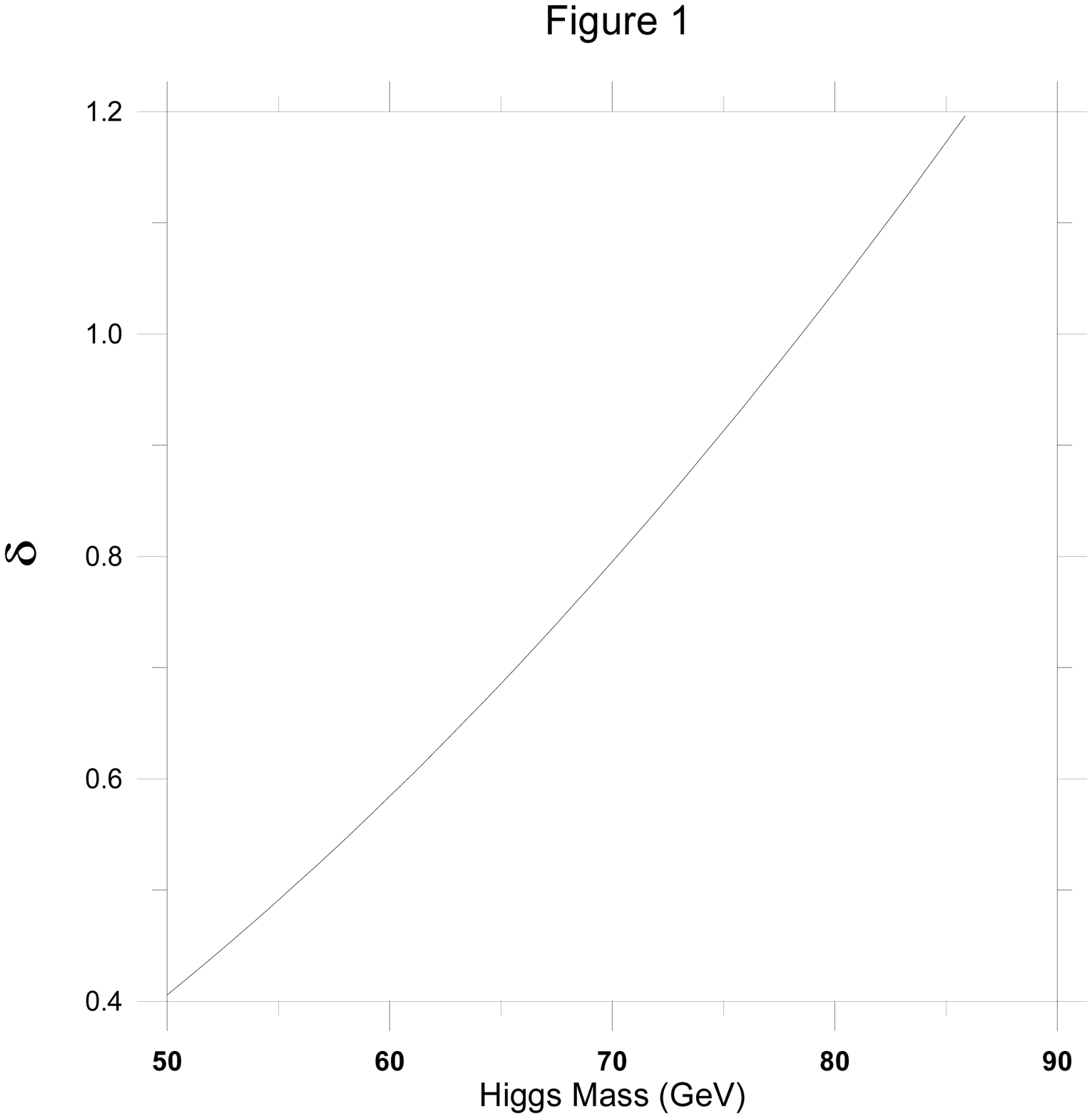}}

\noindent
{\bf Figure 1:} An estimate of the relative magnitude of the leading wave
function contribution, Eq. (\ref{deltaZ}), computed at $T=T_c$.

\epsfysize=20cm
\centerline{\epsfbox{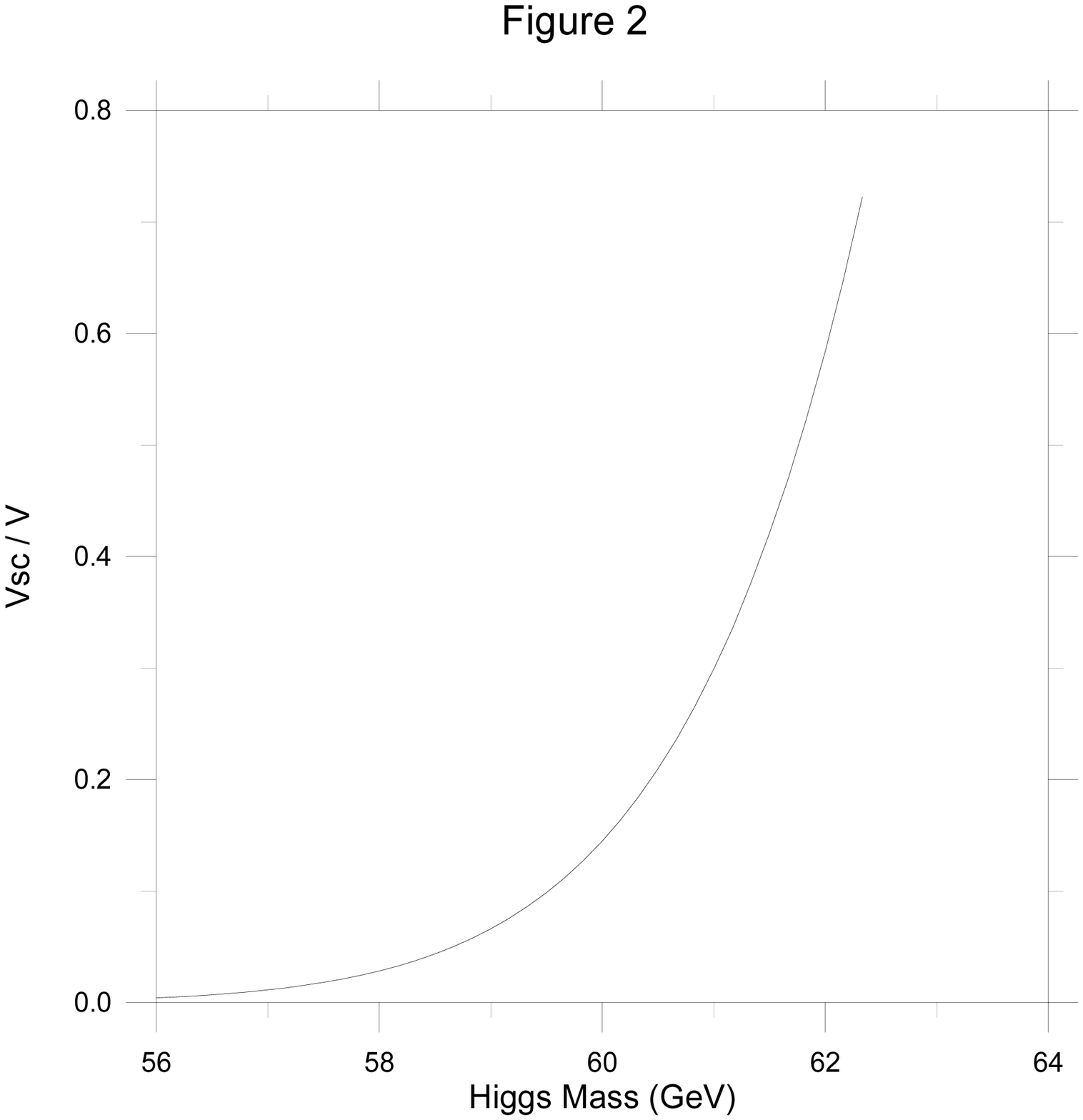}}

\noindent
{\bf Figure 2:} The volume fraction $\frac{V_{\rm sc}}{V}$ of true
vacuum fluctuations (computed at the critical temperature $T_c$)
as a function of the Higgs mass $m_H$ (GeV).

\epsfysize=20cm
\centerline{\epsfbox{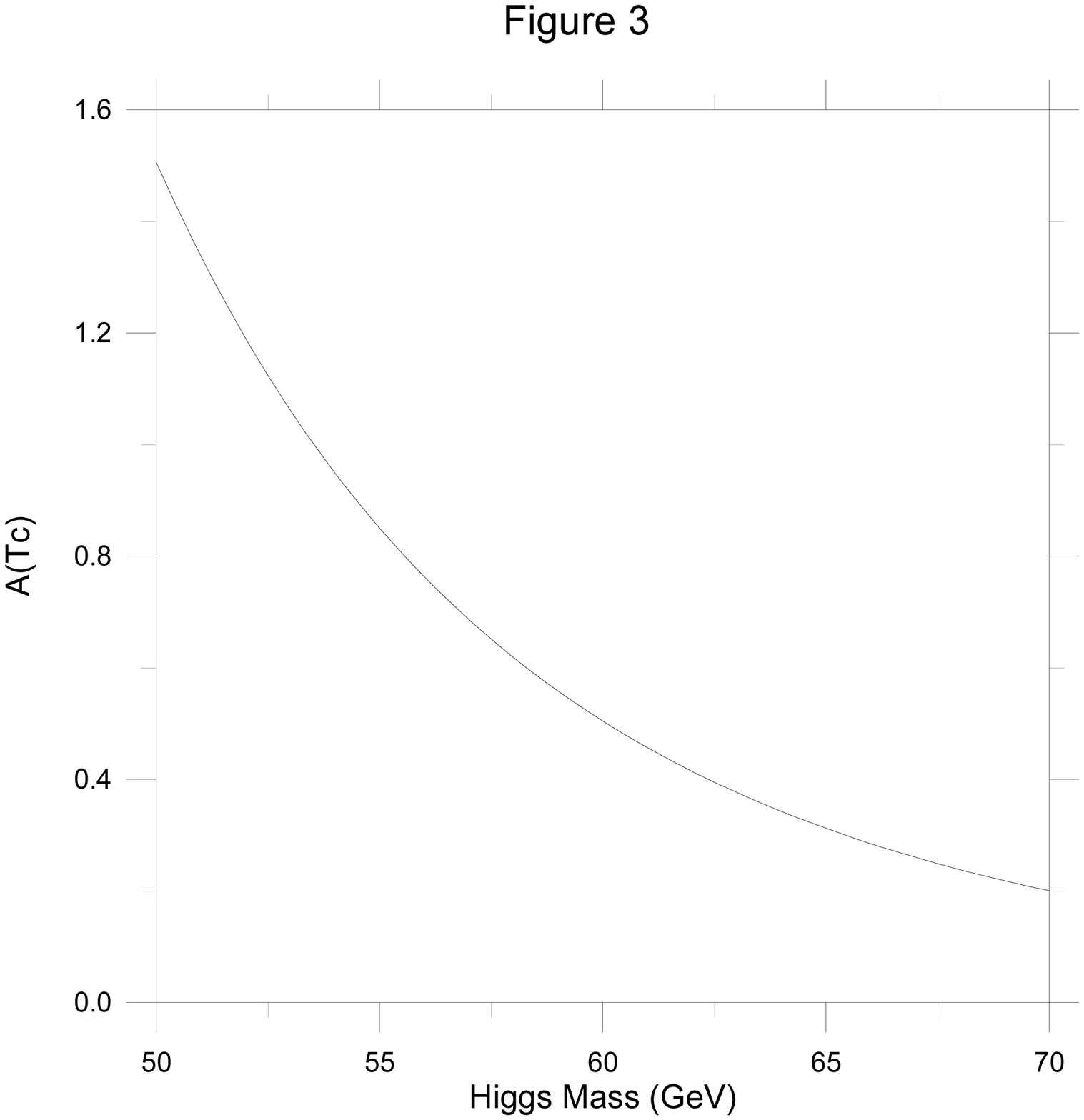}}

\noindent
{\bf Figure 3:} The Prefactor coefficient $A(T)$, computed at the
critical temperature $T_c$, as a function of the Higgs mass.

\epsfysize=20cm
\centerline{\epsfbox{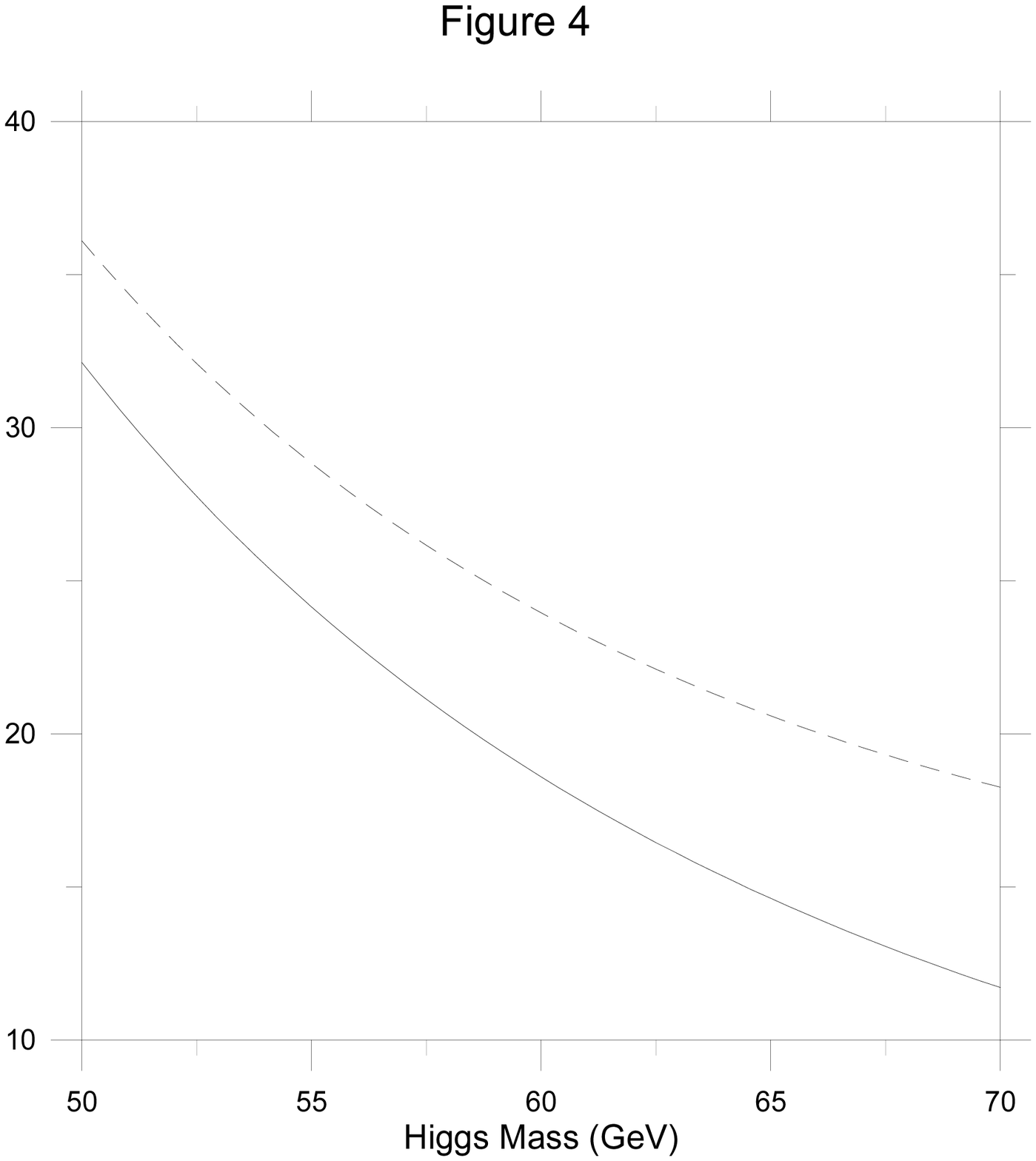}}

\noindent
{\bf Figure 4:} The exponential factor $B(T)=\frac{\Delta F_{\rm eff}
(T)}{T}$ (upper curve)
compared with $F(T)/T$ (lower curve),
both evaluated at $T_c$, as a function of the Higgs mass.

\end{document}